\documentclass[aps,nofootinbib,prd,showpacs,preprintnumbers]{revtex4}%,twocolumn
\usepackage[T1]{fontenc}
\usepackage[utf8]{inputenc}
\usepackage{amsmath}
\usepackage{amsfonts}
\usepackage{amssymb}
\usepackage{amsthm}
\usepackage{caption}
\usepackage{multirow}
\usepackage[english]{babel}
\usepackage{fontenc}
\usepackage{graphicx}
\usepackage{xcolor}
 \usepackage{hyperref}
\usepackage{mathptmx}
\usepackage{caption}
\newcommand {\ignore}[1]{}

\newcommand{\sm}{{standard model }}

\newcommand{\AddrAHEP}{AHEP Group, Institut de F\'{i}sica Corpuscular --
  C.S.I.C./Universitat de Val\`{e}ncia, Parc Cientific de Paterna.\\
  C/Catedratico Jos\'e Beltr\'an, 2 E-46980 Paterna (Val\`{e}ncia) - SPAIN}

\begin{document}

\preprint{IP/BBSR/2017-2}
\title{Probing atmospheric mixing and leptonic CP violation\\
    in current and future long baseline oscillation experiments}
\date{\today}

\author{Sabya Sachi Chatterjee~$^{1,2}$}\email{sabya@iopb.res.in}

\author{Pedro Pasquini~$^3$}\email{pasquini@ifi.unicamp.br}
\author{J.W.F. Valle~$^4$}\email{valle@ific.uv.es, URL:  http://astroparticles.es/}

\affiliation{
$^1$~Institute of Physics, Sachivalaya Marg, Sainik  School Post, Bhubaneswar 751005, India\\
$^2$ Homi Bhabha National Institute, Training School Complex, Anushakti Nagar, Mumbai 400085, India\\
$^3$ Instituto de F\'isica Gleb Wataghin - UNICAMP, {13083-859}, Campinas SP, Brazil\\
$^4$~\AddrAHEP}

\begin{abstract}
  % We perform realistic simulations of the current and future long
  % baseline experiments such as T2K, NO$\nu$A, DUNE and T2HK in order
  % to determine their ultimate potential in probing neutrino
  % oscillation parameters. We focus on the atmospheric angle and the CP
  % phase. We quantify the potential of these experiments to underpin
  % the octant of $\theta_{23}$ as well as the value and sign of
  % $\delta_{CP}$ both in general, as well as within the predictive
  % framework of a previously proposed~\cite{Chen:2015jta} benchmark
  % theory of neutrino oscillations which tightly correlates
  % $\theta_{23}$ and $\delta_{CP}$.

  We perform realistic simulations of the current and future long
  baseline experiments such as T2K, NO$\nu$A, DUNE and T2HK in order
  to determine their ultimate potential in probing neutrino
  oscillation parameters. We quantify the potential of these
  experiments to underpin the octant of the atmospheric angle
  $\theta_{23}$ as well as the value and sign of the CP phase
  $\delta_{CP}$.
  We do this both in general, as well as within the predictive
  framework of a previously proposed~\cite{Chen:2015jta} benchmark
  theory of neutrino oscillations which tightly correlates
  $\theta_{23}$ and $\delta_{CP}$.

\end{abstract}
 \maketitle

 \section{ Preliminaries: a minimal benchmark theory of neutrino
   oscillations}
\label{sec:prel-minim-theory}

The discovery of neutrino oscillations constitutes a major milestone
in particle physics~\cite{Kajita:2016cak,McDonald:2016ixn}.  While
oscillations are a generic expectation in theories of neutrino mass,
the corresponding set of oscillation parameters can be extremely
rich~\cite{Schechter:1980gr}, precluding the possibility of making
detailed predictions for the next generation of oscillation
experiments~\cite{Bandyopadhyay:2007kx}.
Despite the tremendous experimental progress we have had and which
has brought neutrino oscillation physics to the precision age, one
still lacks reliable information, for instance, on the octant of
the atmospheric angle as well as the value of (Dirac-type) CP
phase~\cite{Forero:2014bxa,Capozzi:2016rtj,Esteban:2016qun},
whose determination remains ambiguous.
A generic neutrino oscillation pattern would involve in addition a set
of non-unitarity parameters~\cite{Miranda:2016ptb,Escrihuela:2015wra},
known to bring in a potentially serious ambiguity in probing CP
violation in neutrino oscillations~\cite{Miranda:2016wdr}.

Here we assume the standard three neutrino
paradigm~\cite{Maltoni:2004ei} and perform realistic simulations of
the current and future long baseline oscillation experiments such as
T2K, NO$\nu$A, DUNE and T2HK in order to determine their potential in
probing neutrino oscillation parameters. For definiteness we focus on
the least well-determined ones, namely the atmospheric angle and the
(Dirac-type) CP phase. 

First we quantify the sensitivity of these
experiments to $\theta_{23}$ and $\delta_{CP}$ in general.
We also pose the question within the framework of a simple benchmark
theory of neutrino oscillations proposed in
Ref.~\cite{Chen:2015jta}. Such theory has been proposed from first
principles, based on a warped flavor model naturally predicting light
Dirac neutrinos, so that the lepton mixing matrix has the same
structure as the CKM matrix describing quark mixing. A beautiful
feature of the model consists in the integration of its
extra-dimensional nature, which accounts for the \sm mass hierarchies,
with the implementation of a predictive non-Abelian flavor symmetry,
in our case $\Delta(27)\otimes {\cal Z}_4\otimes {\cal Z}'_4$.
The latter leads to the description of all the four neutrino
oscillation parameters $\theta_{ij}$ and $J_{\rm CP}$, where the
latter is the leptonic CP invariant, in terms of just two angles:
$\theta_\nu$ and $\phi_\nu$ according to the following equations,
 \begin{flalign} 
\sin^2\theta_{12}=&\frac{1}{2-\sin2\theta_\nu\cos\phi_\nu} \nonumber \\%\label{eq:predict2}
\sin^2\theta_{13}=&\frac{1}{3}(1+\sin2\theta_\nu\cos\phi_\nu) \nonumber\\%\label{eq:predict3}
\sin^2\theta_{23}=&\frac{1-\sin2\theta_\nu\sin(\pi/6-\phi_\nu)}{2-\sin2\theta_\nu\cos\phi_\nu} \nonumber\\ 
J_{\rm CP}=&-\frac{1}{6\sqrt{3}}\cos2\theta_\nu
\label{eq:predict4}
\end{flalign}
Given the good determination of $\theta_{13}$ by reactor experiments,
this model is in a sense effectively a one-parameter theory, hence we
call it a ``minimal'' benchmark theory of neutrino oscillations.

Here we explore the potential of current and planned long baseline
oscillation experiments in testing the predictions of this model.
We perform state-of-the-art simulations of the relevant experiments
T2K, NO$\nu$A, DUNE and T2HK in order to ascertain how well they can
probe the model and compare with the situation in a general
unconstrained oscillation scenario.

\section{ Numerical analysis and experimental setups}
\label{simulation}

In order to quantify the sensitivities of the various experimental
setups in testing our benchmark oscillation model, we use GLoBES
\cite{Huber:2004ka,Huber:2007ji} as a numerical simulator. The global
(unconstrained) best fit values of the oscillation parameters in the
three flavor framework, taken from \cite{Forero:2014bxa}, are given
as: $\sin^2\theta_{12}$ = 0.323, $\sin^2\theta_{13}$ = 0.0234,\,
$\sin^2\theta_{23}$ = 0.567 (0.573) for NH (IH) ,\, $\delta_{CP}$ =
1.34$\pi$,\, $\Delta m_{21}^2$ = 7.5$\times10^{-5}\;\rm eV^2$,\,
$\Delta m_{31}^2$ = 2.48$\times10^{-3}$
(-2.38$\times10^{-3}$)\;$\rm eV^2$ for NH (IH).
If specifically not mentioned something else, all the true data have
been generated using the unconstrained best values of the oscillation
parameters. Also, we have considered a fixed hierarchy both in true
and test data.
We are not using any prior on the oscillation
parameters because our test oscillation parameters will be predicted by the model~\cite{Chen:2015jta}. In order to find the
sensitivity of this model at a certain confidence level, we are using
the following Poissionian $\chi^2$ function
\cite{Huber:2002mx,Fogli:2002pt}:
 \begin{eqnarray}
  \chi^2 = \min\limits_{\lbrace\xi_a,\xi_b\rbrace}\left[2\sum_{i=1}^{n}(y_i-x_i-x_i\ln\frac{y_i}{x_i})+\xi^2_a + \xi^2_b \right]
  \label{chi}
 \end{eqnarray}
 where, $n$ is the total number of bins and  
 \begin{eqnarray}
  y_i(\tilde{f}, \xi_a,\xi_b) = N_i^{pre}(\tilde{f})\left[ 1+\pi^a\xi_a \right] + N_i^{b}(\tilde{f})\left[ 1+\pi^b\xi_b \right]
 \end{eqnarray}
 where $\tilde{f}$ denotes the oscillation parameters predicted by the
 model and $\pi^a$,$\pi^b$ denote the systematic errors on signal and
 background respectively, assumed to be uncorrelated between different
 channels. On the other hand $\xi_a$ and $\xi_b$ are the pulls due to
 systematic errors, while $N_i^{pre}$ is the number of predicted
 signal events in the $i$th energy bin and $N_i^{b}$ is the background
 events, where the charged current (CC) background depends on
 $\tilde{f}$. The true data measured by an experiment enter in
 Eq.~\ref{chi} through
 \begin{eqnarray}
  x_i(f) = N_i^{obs}(f) + N_i^{b}(f),
 \end{eqnarray}
 $N_i^{obs}$ is the number of observed CC signal events in the i-th energy bin and $f$ denotes the standard unconstrained oscillation parameters
 whose the best fit values are taken from
 Ref.~\cite{Forero:2014bxa}. Individual contributions coming from the
 various relevant channels are added together in order to get the
 total $\chi^2$ as
 \begin{eqnarray}
  \chi^2_{\rm total} = \underset{\nu_{\mu} \to \nu_e}{\chi^2} + \underset{\bar{\nu}_{\mu} \to \bar{\nu}_e}{\chi^2} +\underset{\nu_{\mu} \to \nu_{\mu}}{\chi^2} + \underset{\bar{\nu}_{\mu} \to \bar{\nu}_{\mu}}{\chi^2}
 \end{eqnarray}

 Finally, this total $\chi^2$ is minimized over the free oscillation
 parameters.
 The simulation runs over four possible experimental scenarios, the
 ``current'' T2K, NOvA experiments and the ``future'' T2HK and DUNE
 proposal setups and this encompass the list of the 
 experiments aimed at improving the $\theta_{23}$ measurements and the
 determination of the CP phase $\delta_{\rm CP}$. For the latter the
 predicted correlation between $\theta_{23}$ and
 $\delta_{\rm CP}$~\cite{Chen:2015jta} can be used to significantly
 shrink down the parameter space of the benchmark model as shown
 in~\cite{Pasquini:2016kwk}. In order to sharpen and extend those
 results we first briefly summarize the experimental setups used in this work.
\begin{enumerate}
\item[1.] {\underline{\bf T2K :}} To simulate the T2K (Tokai to
  Kamiokande) experiment, we assumed the configuration in
  \cite{Abe:2014tzr} with a full exposure of $7.8\times 10^{21}$
  protons on target (POT) which produce an off-axis (angle of $2.5^0$)
  neutrino beam with energy peak around 0.6 GeV hitting a 50 kt
  (fiducial volume 22.5 Kt) water Cerenkov Super-K far detector at
  Kamioka at a distance of 295 km from the the target. In this work,
  half of the total exposure has been assumed in the neutrino mode
  and the remaining half of the exposure in the antineutrino mode. We
  have followed reference \cite{Abe:2014tzr} in great detail,
  reproducing their event spectra in all the modes rather
  well. Following the same reference, we are using an uncorrelated 5\%
  signal normalization error and 10\% background normalization error
  for both neutrino and antineutrino appearance and disappearance
  channels respectively. \\
 
\item[2.]{\underline{\bf T2HK :}} T2HK(Tokai to Hyper-Kamiokande) is
  also a superbeam accelerator based off-axis experiment which is
  expected to be operational around 2025~\cite{Abe:2011ts}. It uses
  the same off-axis setup and the same baseline as T2K. It is supposed
  to be the upgraded version of T2K which also uses a 30 GeV proton
  beam accelerated by the J-PARC facility, which hits the target and
  produces an intense neutrino beam. Following
  Ref.~\cite{Abe:2015zbg}, we assume a 560 kt (fiducial) water Cerenkov far
  detector placed at Hyper-Kamiokande and an integrated beam with power 7.5
  MW$\times 10^7$ sec which corresponds to 1.56$\times10^{22}$ POT. To
  make the event number almost equal for both neutrino and
  antineutrino modes, we have assumed a run time ratio of 1:3 for
  $\nu$:$\bar{\nu}$ that is 2.5 yrs for neutrino mode and 7.5 yrs for antineutrino mode. As a simplified case, we assume an uncorrelated
  5\% signal normalization error and 10\% background normalization
  error for both polarities and for both
  appearance and disappearance channels respectively.\\
 
\item[3.] {\underline{\bf NO$\nu$A :}} NO$\nu$A (NuMI Off-axis $\nu_e$
  Appearance) \cite{Patterson:2012zs,Childress:2013npa} is an off-axis
  accelerator based superbeam experiment, consisting of two detectors,
  one is a near detector at Fermilab and another one is a 14 Kt TASD
  far detector placed in Ash river, Minnesota at an angle $0.8^0$ from
  the beam direction. Neutrinos from NuMI (Neutrinos at the Main
  Injector) will pass through 810 km of earth matter before they are
  detected at the far detector. The off-axis is chosen to get peak
  energy approximately at 2 GeV. NO$\nu$A uses a 120 GeV proton beam
  with beam power 700 kW to produce the intense neutrino beam. The
  expected POT is $3.6\times 10^{21}$ divided in 50\% neutrino mode
  and 50\% anti-neutrino mode, with uncorrelated 5\% signal
  normalization error and 10\% background normalization error for both
  neutrino and antineutrino appearance and disappearance channel
  respectively. All the relevant information has been taken from
  \cite{Agarwalla:2012bv}.\\
  
\item[4.] {\underline{\bf DUNE :}} DUNE is a long baseline future
  generation on-axis superbeam experiment having 1300 km baseline from
  Fermilab to Sanford Underground Research Laboratory in Lead, South
  Dakota. DUNE will use a 40 kt LArTPC as its far detector. We have
  followed the DUNE CDR \cite{Acciarri:2015uup} as reference. It uses
  a 80 GeV proton beam with beam power 1.07 MW with a total exposure
  of 300 kt.MW.yrs having neutrino mode running for 3.5 yrs and
  antineutrino mode running for 3.5 yrs. All other details have been
  matched to the DUNE design report.
  
 \end{enumerate}
 
Before we go to the result section, it is worth to mention that in the numerical simulation we have used a line-averaged constant matter density of 2.8 gm/$\rm{cm}^3$ for T2K, T2HK and NO$\nu$A, and 2.95 gm/$\rm{cm}^3$ for DUNE following the PREM\cite{DZIEWONSKI1981297,stacey:1977} profile.

 \section{Constraining the benchmark model parameters $\theta_{\nu}$ and $\phi_{\nu}$ from experiment}
\label{sec:constr-thet-phi_nu}

\begin{figure}[h!]
 \includegraphics[height=7cm,width=7cm]{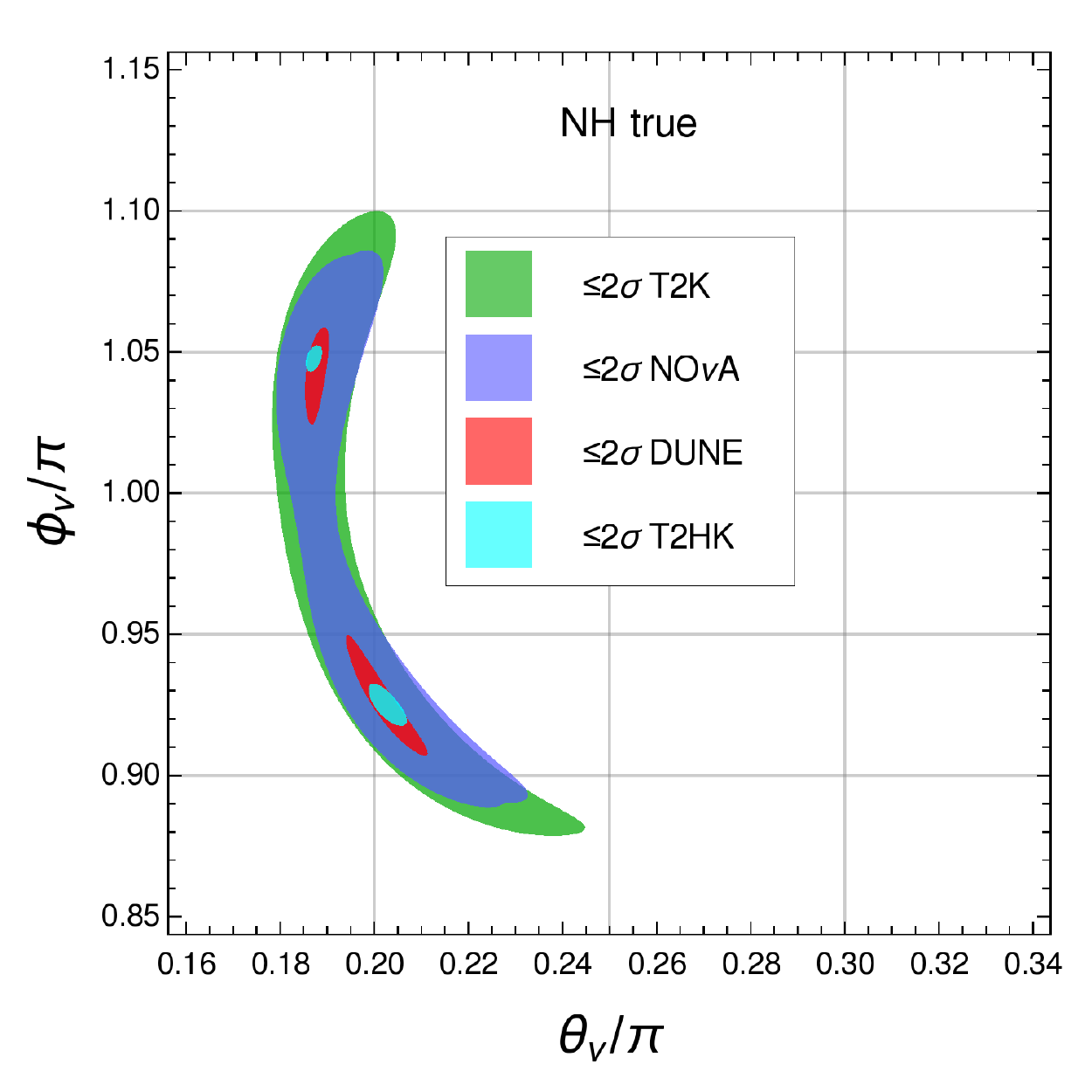}
 \includegraphics[height=7cm,width=7cm]{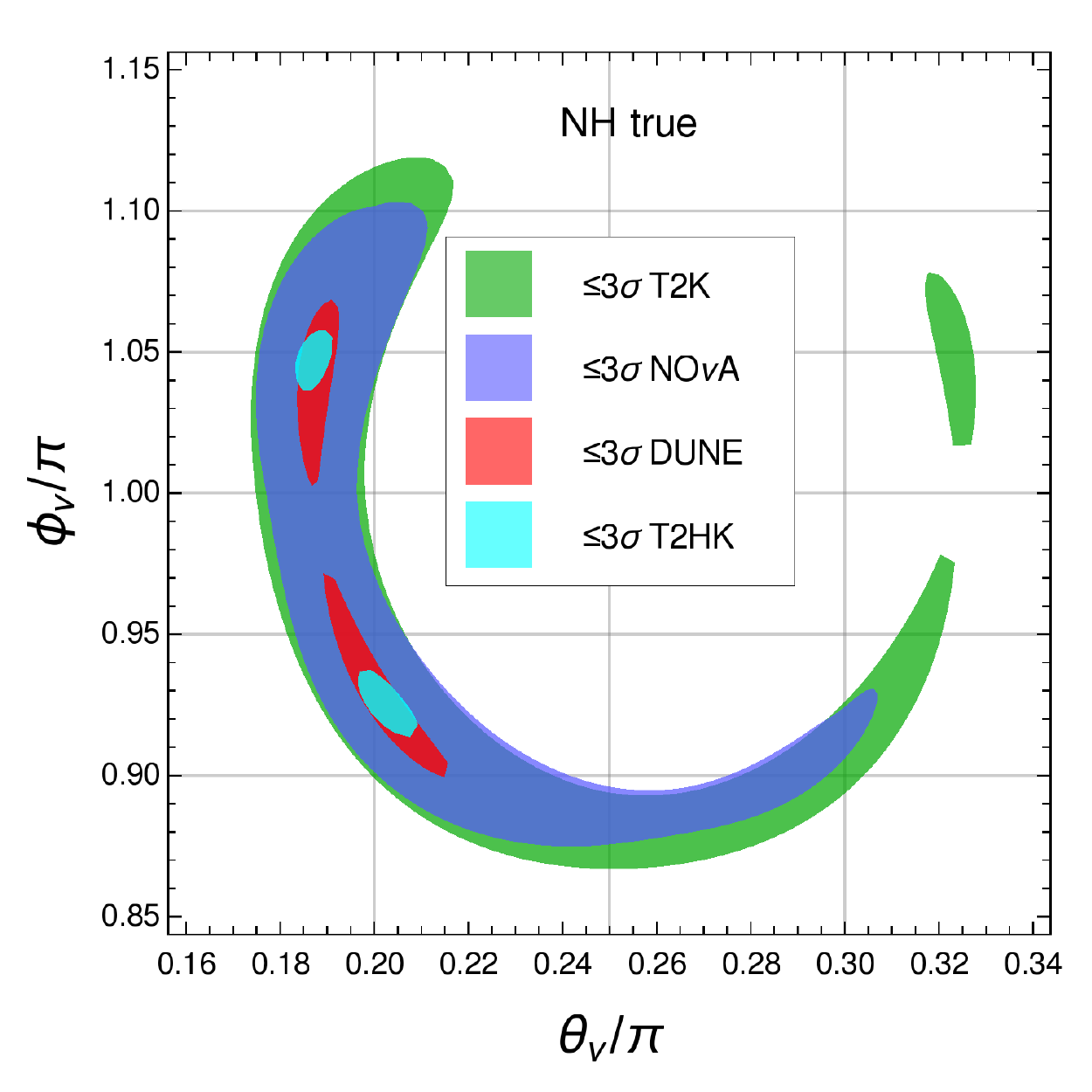}\\
 \caption{Allowed regions of the two model parameters $\theta_{\nu}$
   and $\phi_{\nu}$ at 2$\sigma$ (left) and 3$\sigma$ (right)
   confidence level at 1 d.o.f. that is ($\Delta\chi^2$ = 4, 9 respectively). The plots assume Normal Hierachy (NH)
   as true.  The dark green band represents the sensitivity of T2K,
   while the blue band corresponds to NO$\nu$A.  The red and cyan
   bands give the expected sensitivities of the DUNE and T2HK
   experiments.}
  \label{region1}
 \end{figure}

 Equations \ref{eq:predict4}, expressed in terms of two free
 parameters $\theta_{\nu}$ and $\phi_{\nu}$ suggest that our benchmark
 model can be tested directly in low energy long baseline (LBL)
 neutrino oscillation experiments by obtaining the oscillation
 probability as a function of these two parameters and comparing to
 experimental data. This will lead to a restriction at a certain
 confidence level. In this section, we present the allowed region of
 the two model parameters $\theta_{\nu}$ and $\phi_{\nu}$ implied by
 the current and future LBL experiments.

%%  ggg

 Figure \ref{region1} represents the restricted region of the two 
 parameters $\theta_{\nu}$ and $\phi_{\nu}$ at 2$\sigma$ (left panel)
 and 3$\sigma$ (right panel) confidence level at 1 degree of freedom
 assuming normal hierarchy (NH) as our true choice. The dark
 green band represents the allowed region given by T2K, the blue band
 is obtained from NO$\nu$A, the red band is the sensitivity region
 expected for DUNE and the Cyan band corresponds to the sensitivity
 region of the proposed T2HK experiment.
 True data set has been generated using the unconstrained values of
 the oscillation parameters as mentioned in sec.~\ref{simulation} and
 then fitted to the test data set obtained from each pair of
 $\theta_{\nu}$ and $\phi_{\nu}$ in order to calculate the minimum
 $\Delta\chi^2$. Now the same procedure has been followed for all allowed \footnote{As pointed out by~\cite{Chen:2015jta}, the model allows both NH
 (for $\theta_\nu\in[0,\pi/2] \cup [3\pi/2,2\pi]$) and IH
 ($\theta_\nu\in[\pi/2,3\pi/2]$). For definiteness here we consider
 only NH in the region $\theta_\nu\in[0,\pi/2]$. The angle $\phi_\nu$ can assume any value in between 0 to 2$\pi$. } values
 of $\theta_{\nu}$ and $\phi_{\nu}$.
 %varied within their allowed region 0 to 2$\pi$. 
%
 In order to obtain these sensitivity bands, we only consider those
 values of the new parameters for which model can be tested at certain
 confidence level that is $\Delta\chi^2\leq$ n$\sigma$ (here, n = 2,
 3). 

 From Fig.\ref{region1}, it is quite evident that the T2HK experiment
 is expected to provide the best sensitivity on the  model parameters, followed by DUNE.
 The performance of T2HK is best because of low baseline and huge
 statistics which implies a very precise measurement of $\delta_{CP}$,
 an essential ingredient to constrain our reference benchmark
 model. Note that for DUNE, the CP sensitivity is somewhat less than T2HK.
 On the other hand NO$\nu$A gives somewhat better sensitivity than T2K.

 In table \ref{tab:predic1}, we show a fair comparison between the
 model independent (unconstrained) oscillation parameters and the one
 predicted by our simple benchmark model in different experiments. The
 minimum value of the $\Delta\chi^2$ coming from different experiments
 is also shown within parenthesis for the corresponding experiment.
 One should keep in mind that this analysis assumes that the {\it true
   values} is the minimum of the current global neutrino oscillation
 fit. Since the latter assumes the unconstrained scenario with its 4
 free parameters, it follows that the true values in the simulation
 cannot be reproduced by the our benchmark model which has only 2
 parameters, lying $2\sigma$ away from the
 minimum~\cite{Chen:2015jta}.
\begin{table}[h!]
  \centering
  \begin{tabular}{cccccc}
  \hline \hline 
  Parameter & DUNE ($\chi^2_{min}$ = 0.14)& T2HK ($\chi^2_{min}$ = 0.637) & NO$\nu$A ($\chi^2_{min}$ = 0.016)& T2K ($\chi^2_{min}$ = 0.015) & Unconstrained case\\ \hline \hline
  $s_{12}^2$&  0.341 & 0.341 & 0.341 &0.341 & 0.323($\pm$0.016)\\
  $s_{13}^2$ & 0.023 & 0.023 & 0.024 &0.024 & 0.0234($\pm$0.0020) \\
  $s_{23}^2$ & 0.567 & 0.565 &0.565 &0.566 & 0.567($\substack{+0.025 \\ -0.043}$)\\
  $\delta_{\rm CP}/\pi$  & 1.30 & 1.30 & 1.30 & 1.30 &  1.34($\substack{+0.64 \\ -0.38}$)\\ \hline
  \end{tabular}
  \caption{Values of the neutrino oscillation parameters corresponding
    to the $\chi^2$ minima obtained from the benchmark model.
    % $\theta_\nu=0.295\pi$ and $\phi_\nu=0.92\pi$,
    The sixth column denotes the standard ``unconstrained''
    three-neutrino best fit values for NH taken
    from~\cite{Forero:2014bxa}.  The number within the parenthesis
    indicates the minimum value of the $\chi^2$ predicted from the
    benchmark model for the corresponding experiment.  }
    \label{tab:predic1}
\end{table}  

 \section{Sensitivities on oscillation parameters  }
\label{sec:sens-oscill-param}
 
 Here we examine the sensitivities on neutrino mixing parameters and
 CP phase, specially focusing to $\theta_{23}$ and $\delta_{CP}$, currently the
 two most poorly determined oscillation parameters.
 Before presenting our results notice that oscillation studies can be
 used to probe oscillation parameters in two ways: either in the
 general unconstrained three-neutrino scenario or within the above
 minimal benchmark picture of neutrino oscillations.
 In other words, by assuming the general oscillation picture as the
 truth, we expect that our available oscillation parameter space will
 be highly restricted by future experiments in the benchmark scenario.
 Alternatively, by taking our minimal benchmark picture as true, the
 real minimum of the oscillation parameters differs from the one
 obtained by the global oscillation fit, which assumes general
 $\chi^2$ minimization with four free parameters.
These two possible interpretations require a careful analysis. In
order to do that one should analyze and compare both schemes in the
same footing for each experiment.

 \begin{figure}[h!]
\centering
 \includegraphics[height=7cm,width=7cm]{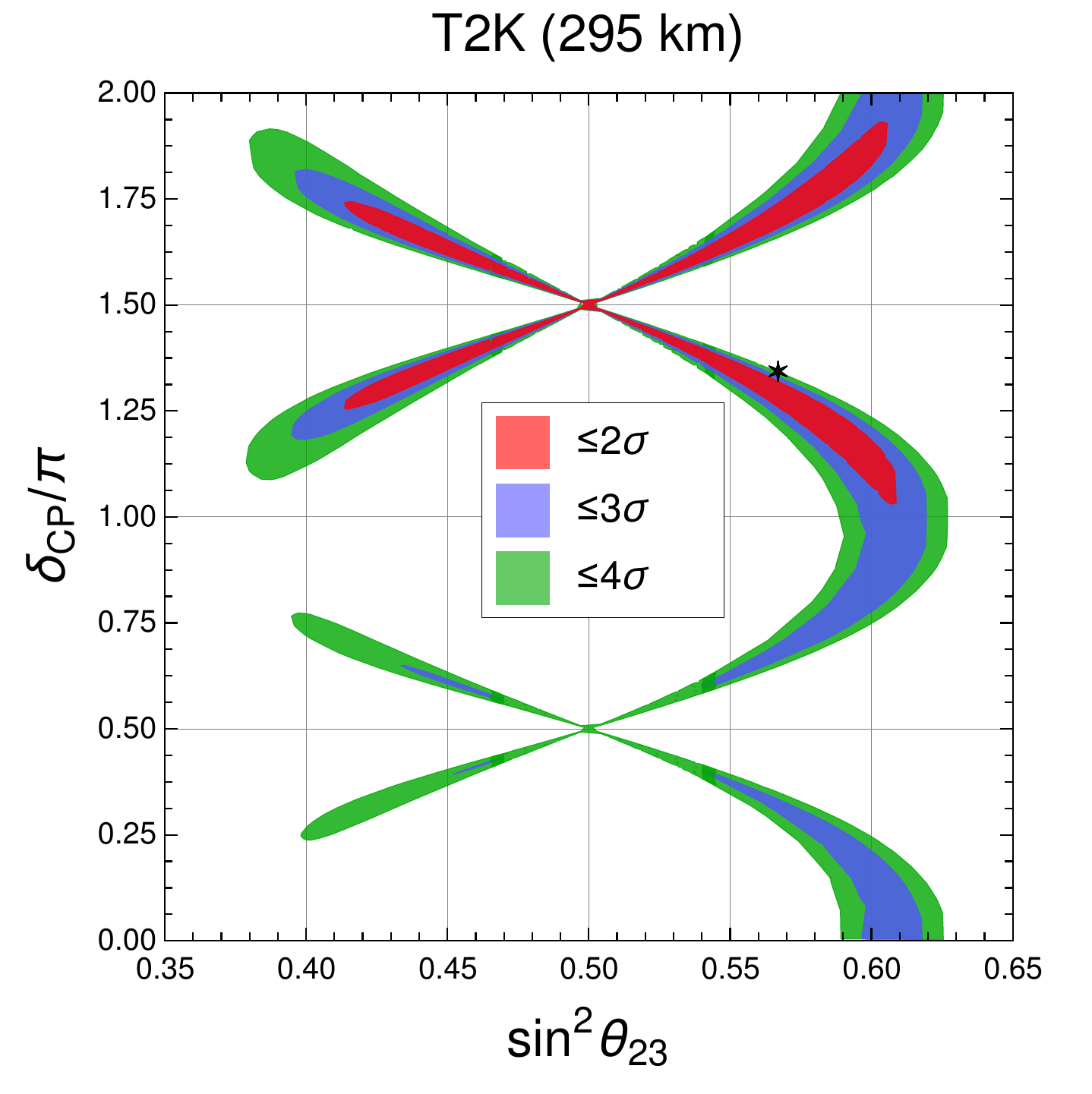}
 \includegraphics[height=7cm,width=7cm]{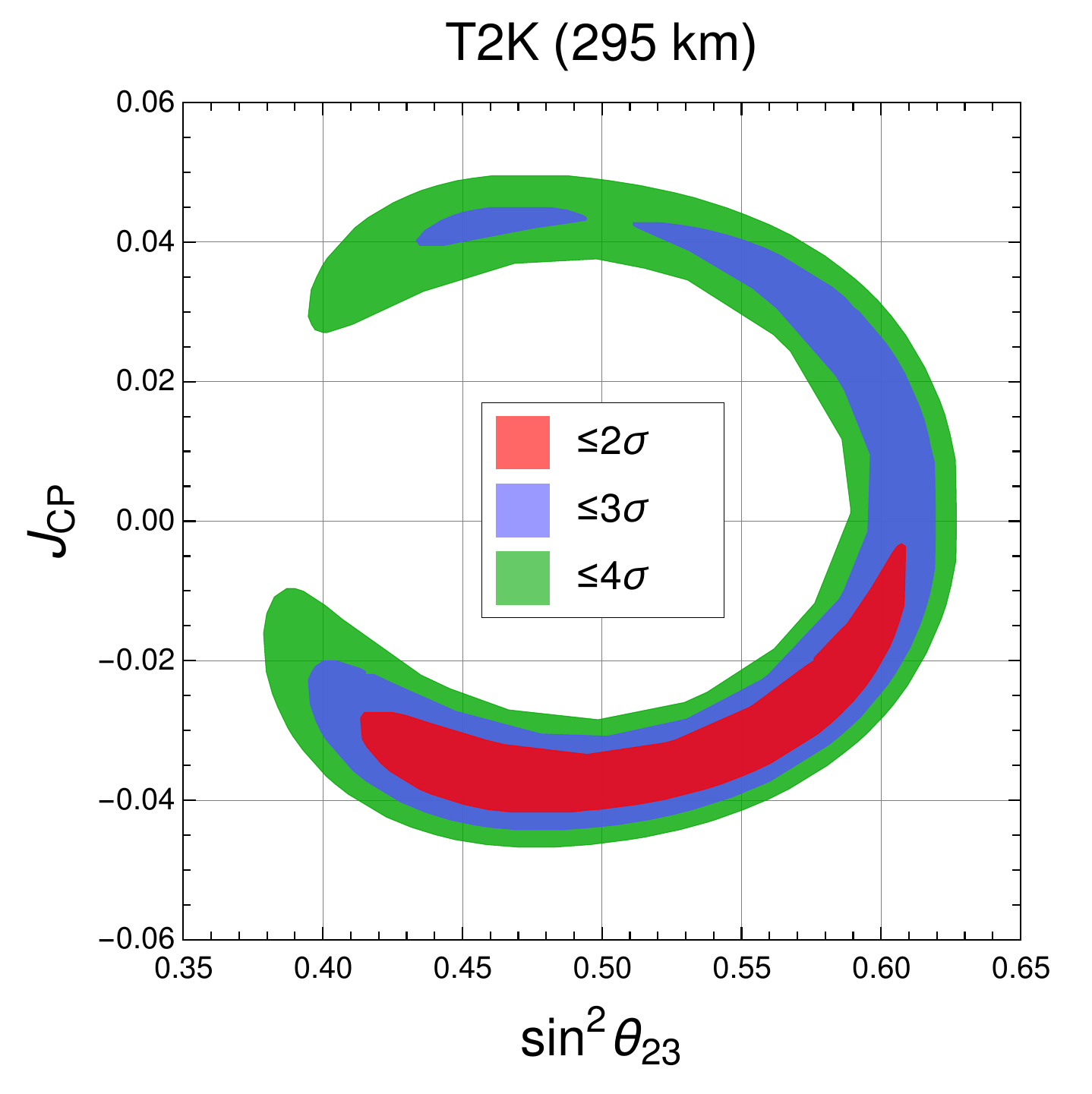}\\
 \includegraphics[height=7cm,width=7cm]{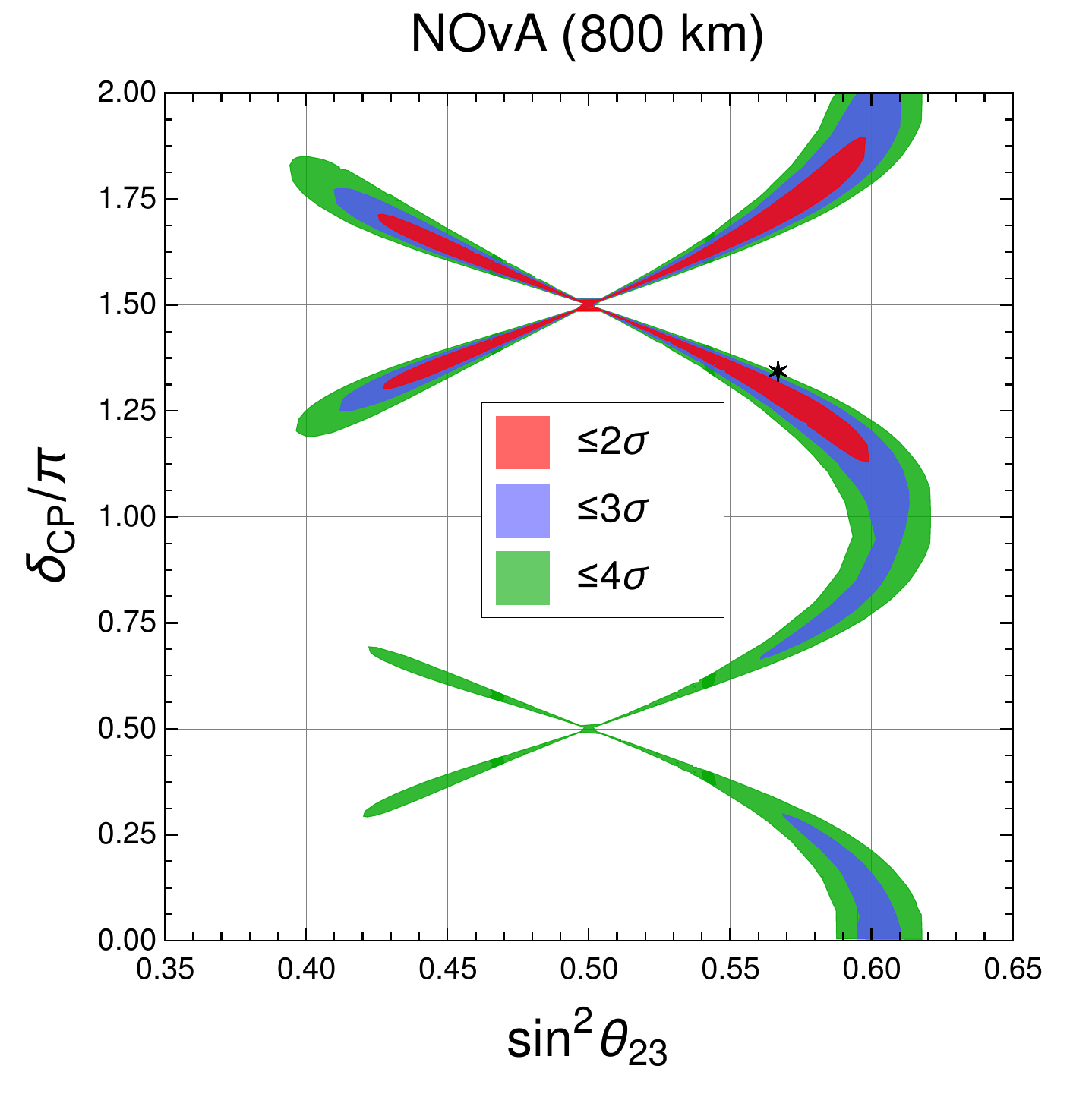}
 \includegraphics[height=7cm,width=7cm]{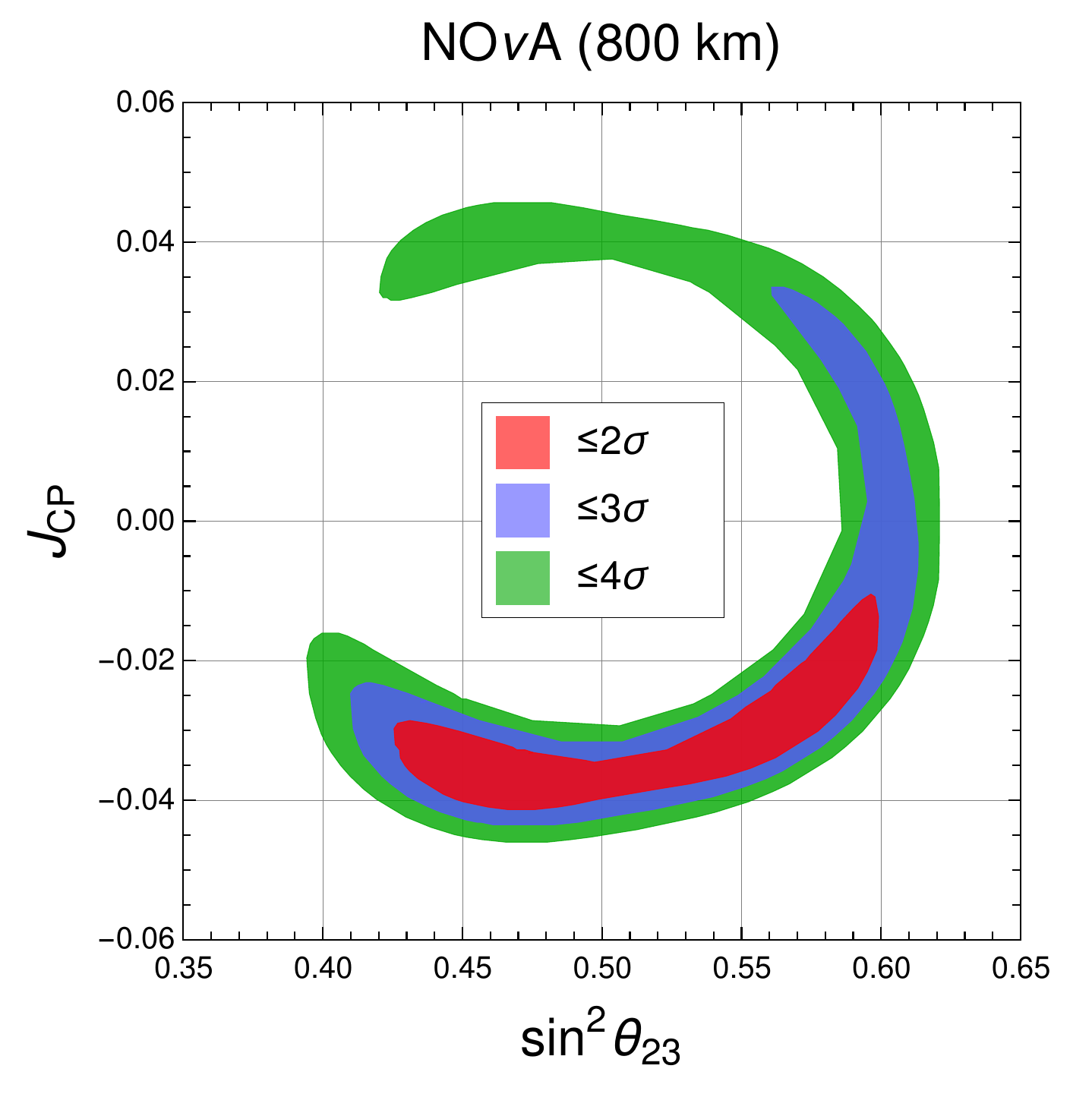}
 \\
 \caption{Precision ``measurement'' of $\sin^2\theta_{23}$ and
   $\delta_{CP}$ at T2K and NO$\nu$A as predicted by the benchmark
   model when NH is the true hierarchy. The star denotes the
   unconstrained values from the fifth column of table
   \ref{tab:predic1} and the bands correspond to the 2$\sigma$,
   3$\sigma$, and 4$\sigma$ C.L uncertainties.  }
  \label{region2}
 \end{figure}

\subsection{Sensitivity of T2K and NO$\nu$A to $\theta_{23}$ and $\delta_{CP}$  in the  minimal benchmark oscillation model}
\label{sec:prob-thet-delt-1}

  The results from Section~\ref{sec:constr-thet-phi_nu} can be
  translated from the two parameters of our benchmark model into the
  four free parameters $\theta_{ij}$ and $\delta_{\rm CP}$ describing
  oscillations, through Eq.~\ref{eq:predict4}, obtaining a
  $\chi^2_{\rm 0}$,
\begin{equation}
\chi^2_{\rm 0}\equiv \chi^2(\theta_{ij}(\theta_\nu,\phi_\nu),\delta_{\rm CP}(\theta_\nu,\phi_\nu))
\end{equation}
which is the $\chi^2$ function relevant if one assumes the standard
picture as true.  For definiteness we assume NH to be the true
hierarchy. The corresponding two-dimensional $2,3$ and $4\sigma$
contours for the T2K and NO$\nu$A experiments are presented in
Fig.\ref{region2}. These are the values of the parameters
$\theta_{23}$ and $\delta_{CP}$ which actually contribute to delimit
the bands indicated in Fig.\ref{region1}.
The left panels give the $\sin^2\theta_{23}$ vs $\delta_{CP}$ contour
plot, while the right panels are the $\sin^2\theta_{23}$ versus
$J_{CP}$ contour plots, where $J_{CP}$ is the CP invariant.  The upper
(lower) panels of Fig.\ref{region2} correspond to T2K (NO$\nu$A).  The
red band in each plot of Fig.\ref{region2} corresponds to the
2$\sigma$ C.L. allowed region, the blue band corresponds to 3$\sigma$
C.L. and the green corresponds to the 4$\sigma$ C.L. allowed region. The star
denotes the unconstrained values taken from the fifth column of
table \ref{tab:predic1}.

Notice the clear correlation between $\theta_{23}$ and $\delta_{CP}$
which is a consequence of Fig.~\ref{region1}. Note also, that a
maximal choice of $\theta_{23}$ corresponds to the maximal CP
violation (up to sign) for T2K and NO$\nu$A which is a very important
prediction of the benchmark model.  Moreover, for non-maximal values
of $\theta_{23}$, there is a four fold degeneracy in the CP phase
determination in T2K and NO$\nu$A. Apart from the $\theta_{23}$ -
$\delta_{CP}$ four-fold degeneracy, there is also degeneracy between
the lower octant ($\sin^2\theta_{23}$ < 0.5) and higher octant
($\sin^2\theta_{23}$> 0.5), so that, this two parameter model can not
distinguish the octant of the atmospheric angle $\theta_{23}$. As
expected, in the $J_{CP}$ plots the degeneracy is clearly reduced.

\subsection{Sensitivity of T2K and NO$\nu$A to $\theta_{23}$ and $\delta_{CP}$  in the general 3-neutrino oscillation picture}
\label{sec:prob-thet-delt}

\begin{figure}[h!]
\centering
 \includegraphics[height=7cm,width=7cm]{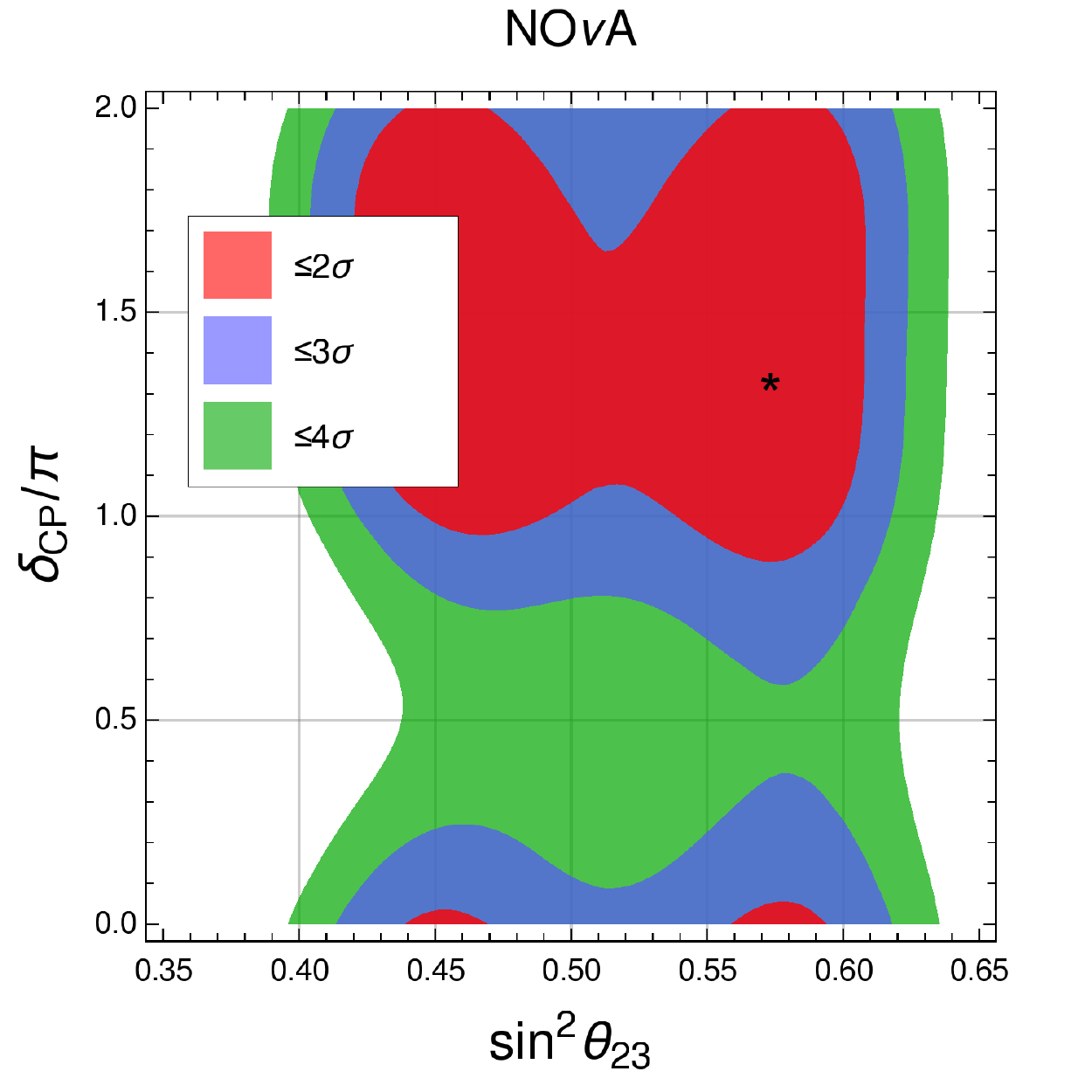}
 \includegraphics[height=7cm,width=7cm]{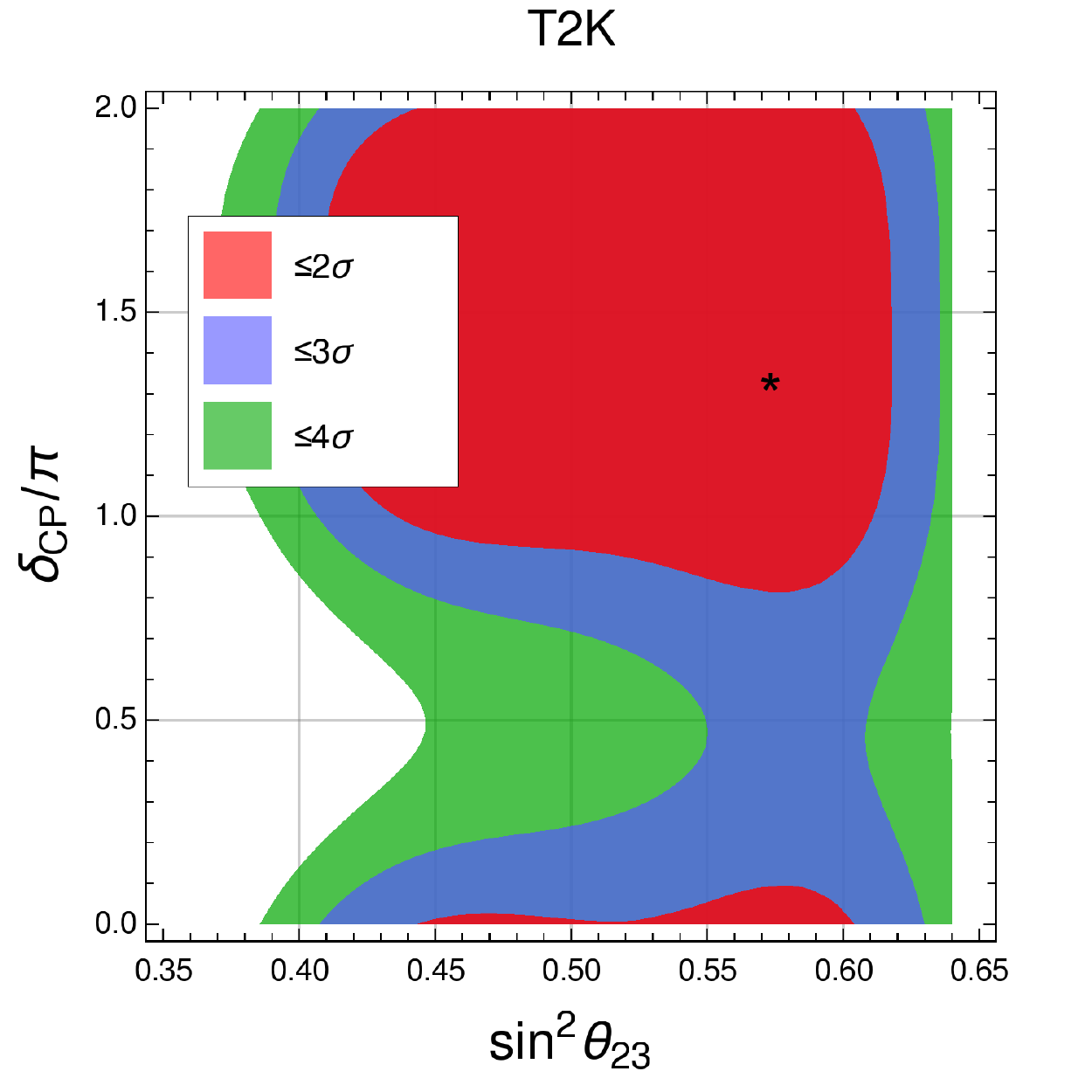}
 \caption{Precision ``measurement'' of $\sin^2\theta_{23}$ and
   $\delta_{CP}$ at T2K and NO$\nu$A for generic unconstrained
   3-neutrino oscillations when NH is the true hierarchy. The star
   denotes the unconstrained values taken from the fifth column of
   table \ref{tab:predic1} and the bands correspond to the 2$\sigma$,
   3$\sigma$, and 4$\sigma$ C.L uncertainties.  }
  \label{precision-t2k}
  \end{figure}

  Here we summarize our model independent results for the oscillation
  parameters $\theta_{23}$ and $\delta_{CP}$. They hold in the general
  3-neutrino oscillation picture assuming again NH to be the true
  hierarchy.  The precision ``measurements'' of the oscillation
  parameters $\sin^2\theta_{23}$ and $\delta_{CP}$ in the T2K and
  NO$\nu$A experiments are given in Fig.~\ref{precision-t2k}. The star
  symbol corresponds to the unconstrained Global best-fit values of
  the oscillation parameters as given in table~\ref{tab:predic1}. The
  red, blue and dark green bands in each plot correspond to the
  2$\sigma$, 3$\sigma$ and 4$\sigma$ uncertainties respectively in
  $\sin^2\theta_{23}$ and $\delta_{CP}$
  plane. Fig.~\ref{precision-t2k} clearly reflects the physics
  potential of T2K and NO$\nu$A in reconstructing the CP phase
  $\delta_{CP}$ and atmospheric mixing angle $\theta_{23}$
  corresponding to the point denoted by the symbol "star".  Even if
  for a fixed phase, there is a degeneracy between the two octants
  of the atmospheric angle  $\theta_{23}$ at 2$\sigma$ C.L. for both 
  experiments.
  
  Notice that the unconstrained best fit does
  not coincide with the minimum predicted by the model because the
  true value cannot be reproduced perfectly within the model. This
  implies that our benchmark oscillation scheme finds different
  minimum values for the current/expected oscillation parameters 
  than obtained in an unconstrained fit.
%
%   In each plot the green band corresponds to the 2$\sigma$ C.L
%   uncertainty, the blue band corresponds to the 3$\sigma$ C.L
%   uncertainty, and the red band corresponds to the 4$\sigma$ C.L
%   uncertainty in the parameter space.
%

\begin{figure}[h!]
\centering
 \includegraphics[height=7cm,width=7cm]{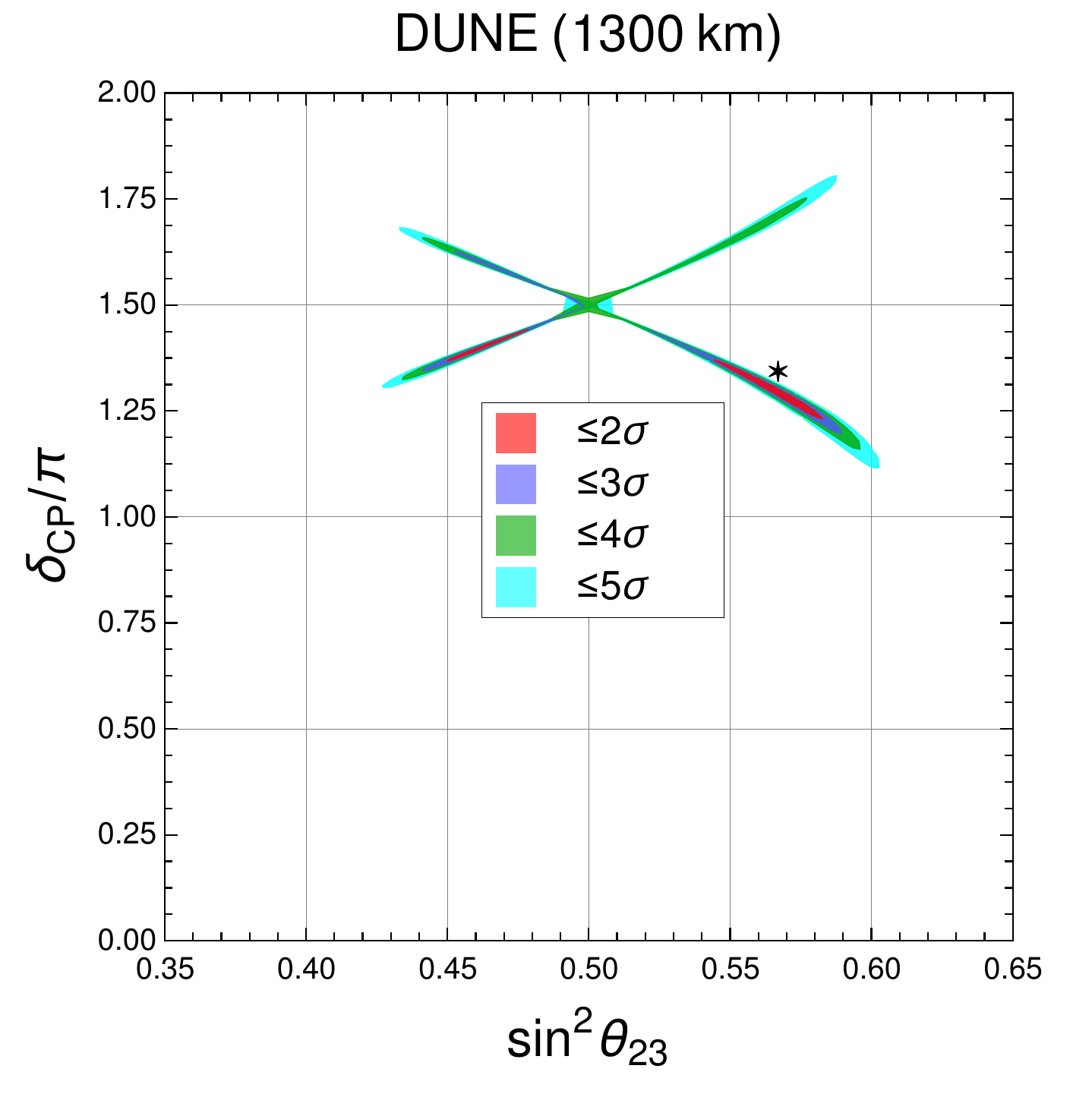}
 \includegraphics[height=7cm,width=7cm]{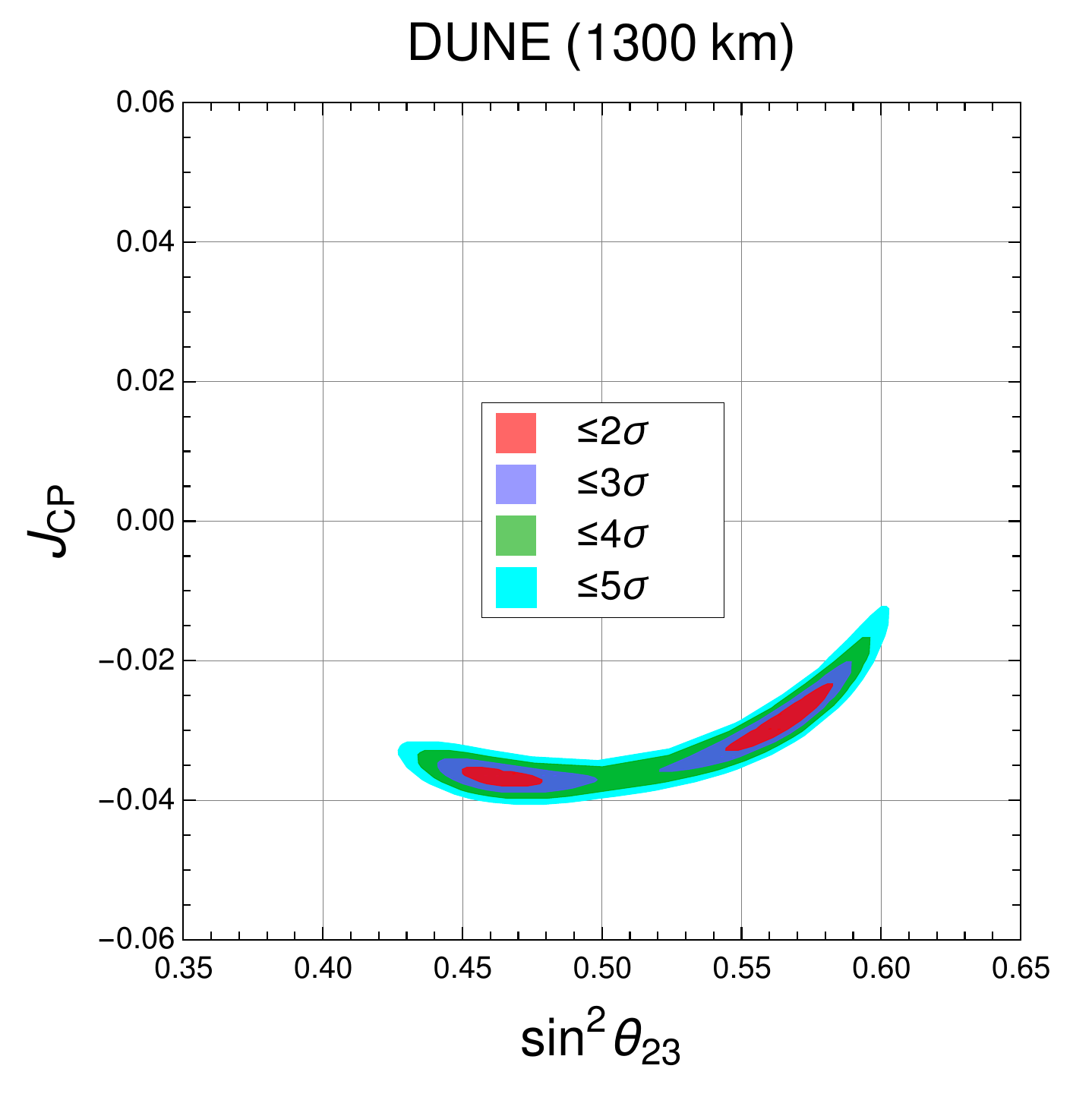}
\includegraphics[height=7cm,width=7cm]{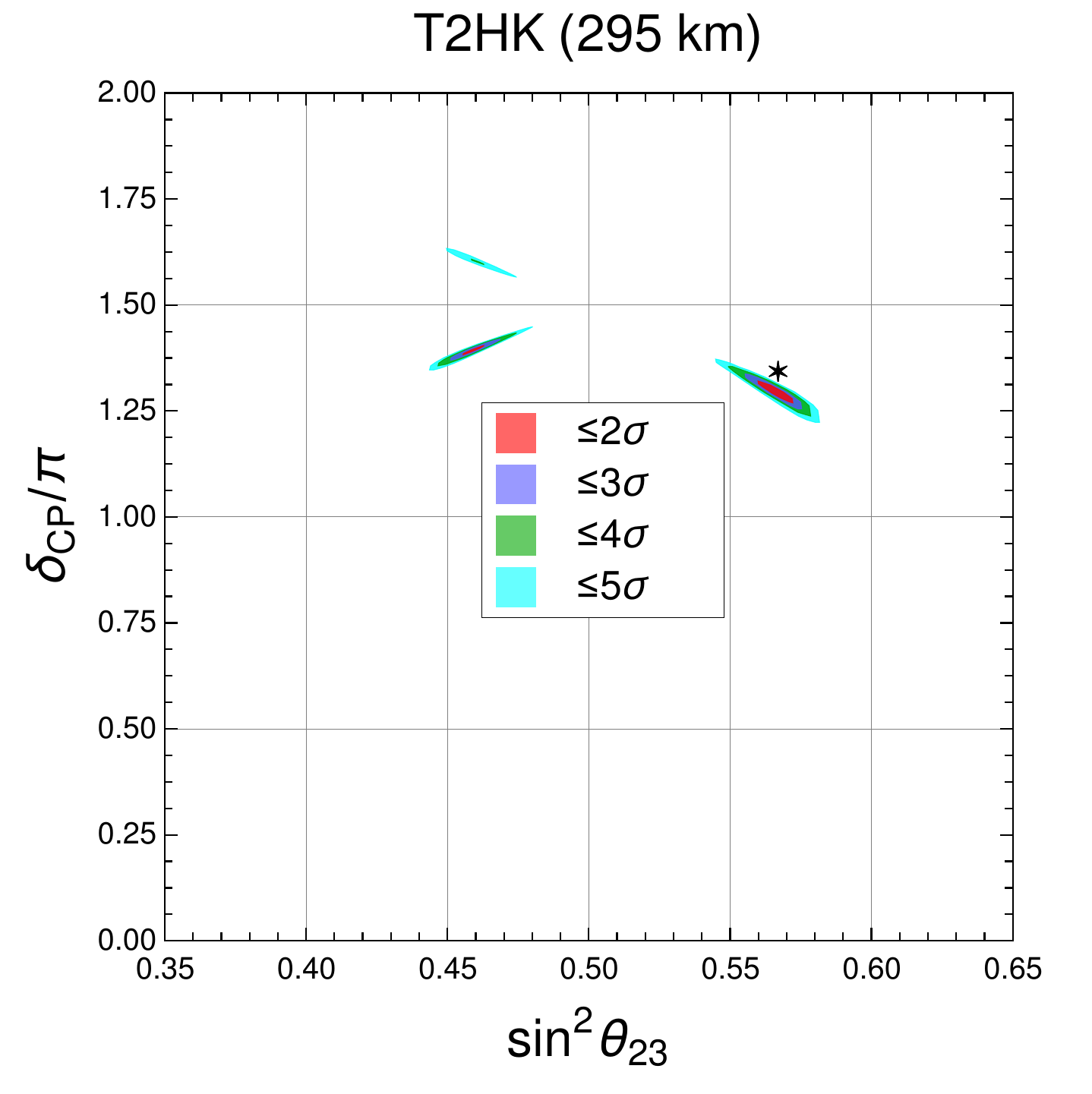}
 \includegraphics[height=7cm,width=7cm]{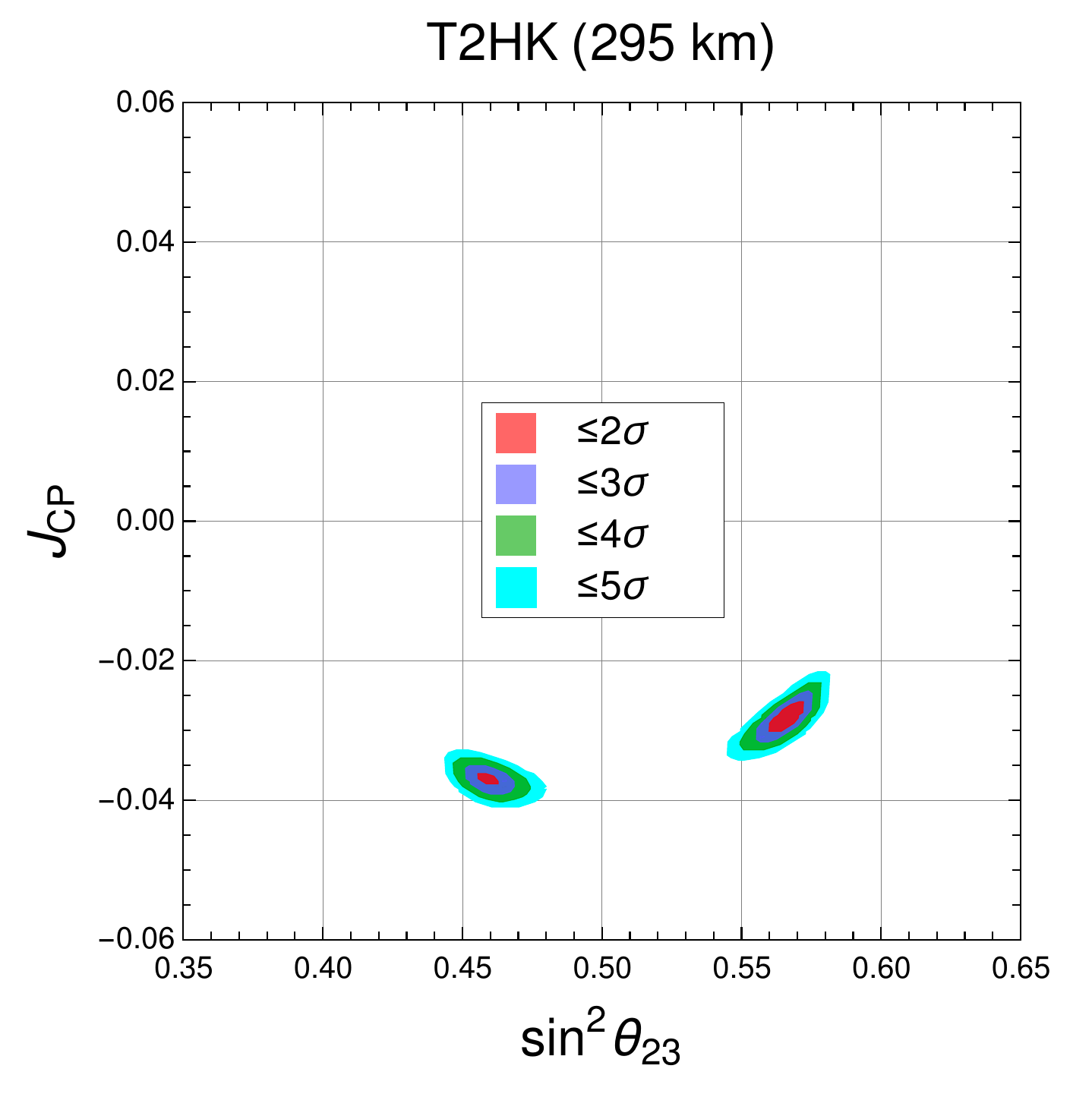}
 \caption{Precision ``measurement'' of $\sin^2\theta_{23}$ and
   $\delta_{CP}$ at future LBL experiments DUNE and T2HK when NH is
   the true hierarchy.  The star denotes the unconstrained values
   taken from the fifth column of table \ref{tab:predic1}. The bands
   correspond to the 2,~3,~4 and 5$\sigma$ C.L uncertainty. }
  \label{octant_sensitivity1}
  \end{figure}

\subsection{Sensitivity of future experiments  }
\label{sec:probing-model-future}

\begin{figure}[h!]
\centering
\includegraphics[height=7cm,width=7cm]{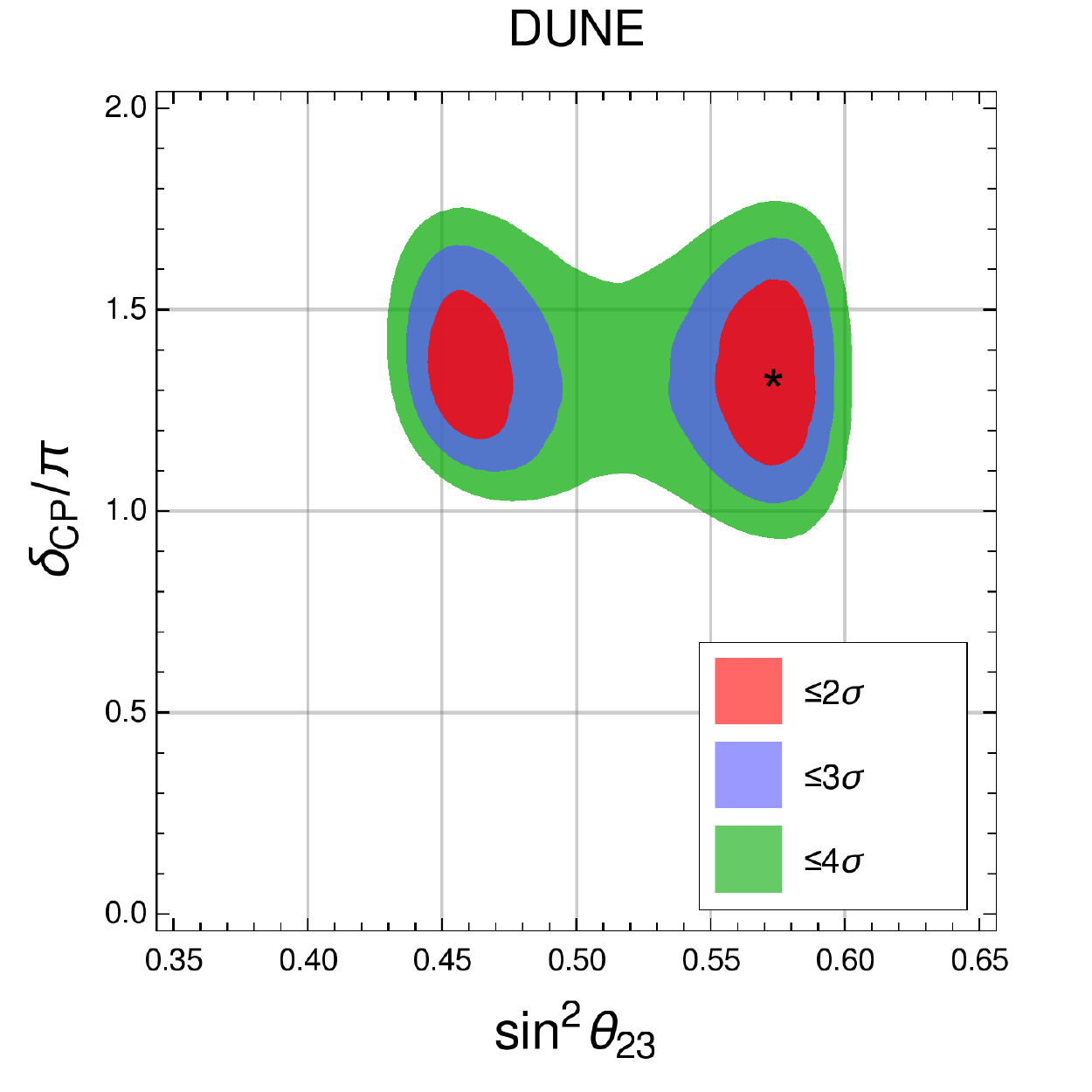}
 \includegraphics[height=7cm,width=7cm]{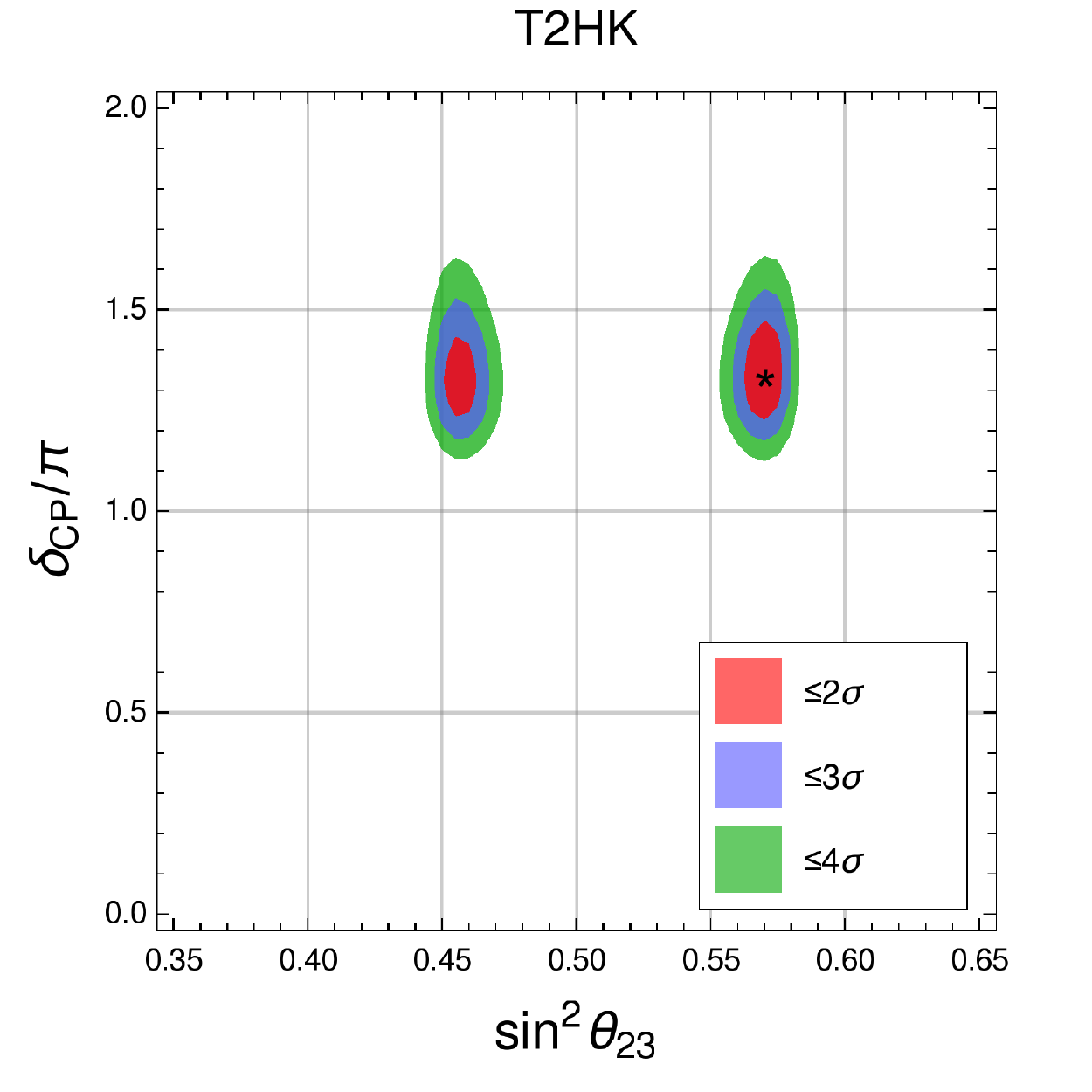}
\caption{Precision ``measurement'' of $\sin^2\theta_{23}$ and
  $\delta_{CP}$ for generic unconstrained 3-neutrino oscillations when
  NH is the true hierarchy. The star denotes the unconstrained values
  taken from the fifth column of table \ref{tab:predic1}. The bands
  correspond to the 2$\sigma$, 3$\sigma$ and 4$\sigma$ C.L
  uncertainty. Notice that in this case the octant would remain unresolved even at 2$\sigma$ C.L.  }
  \label{precision}
  \end{figure}
 \begin{figure}[h!]
  \includegraphics[height=7cm,width=7cm]{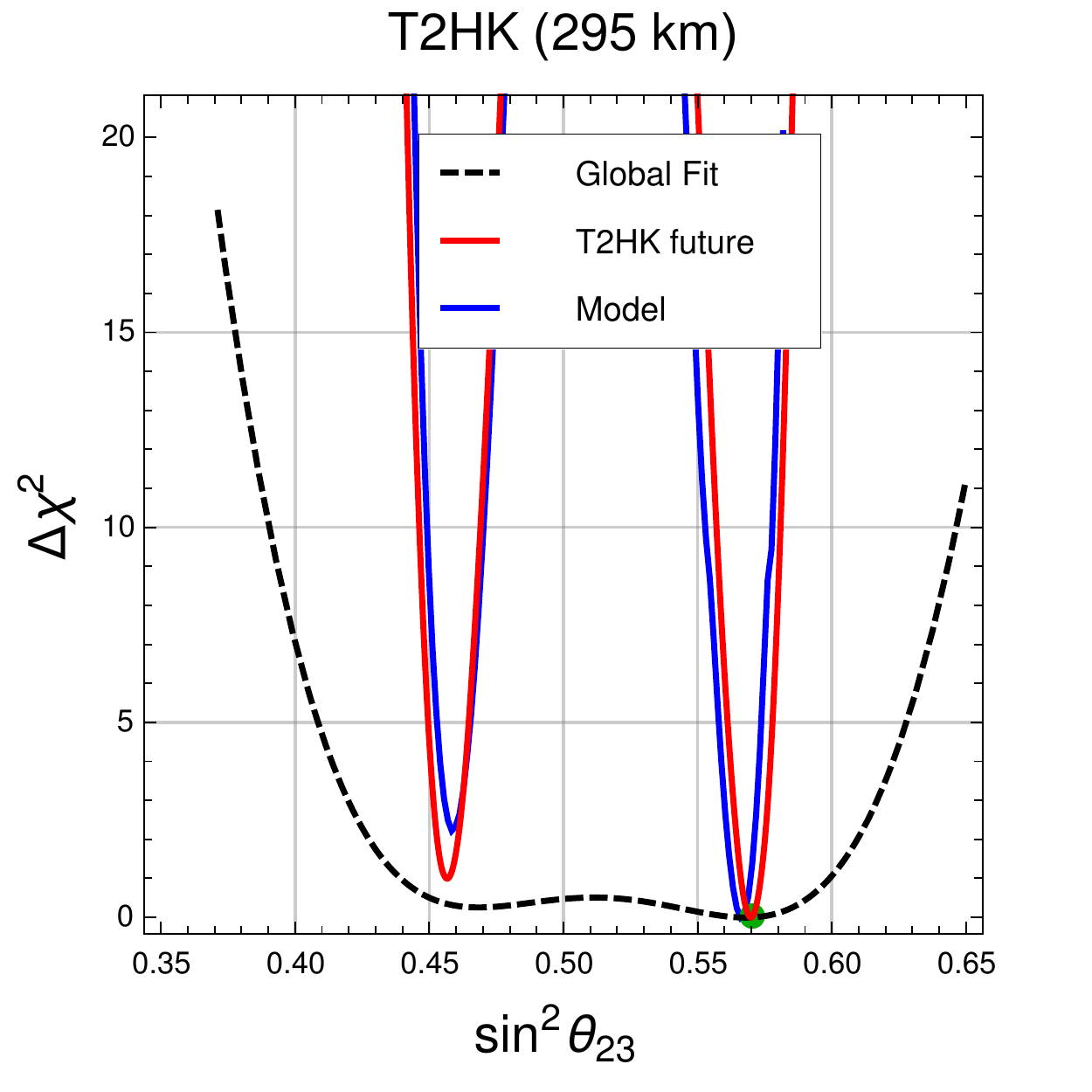}
  \includegraphics[height=7cm,width=7cm]{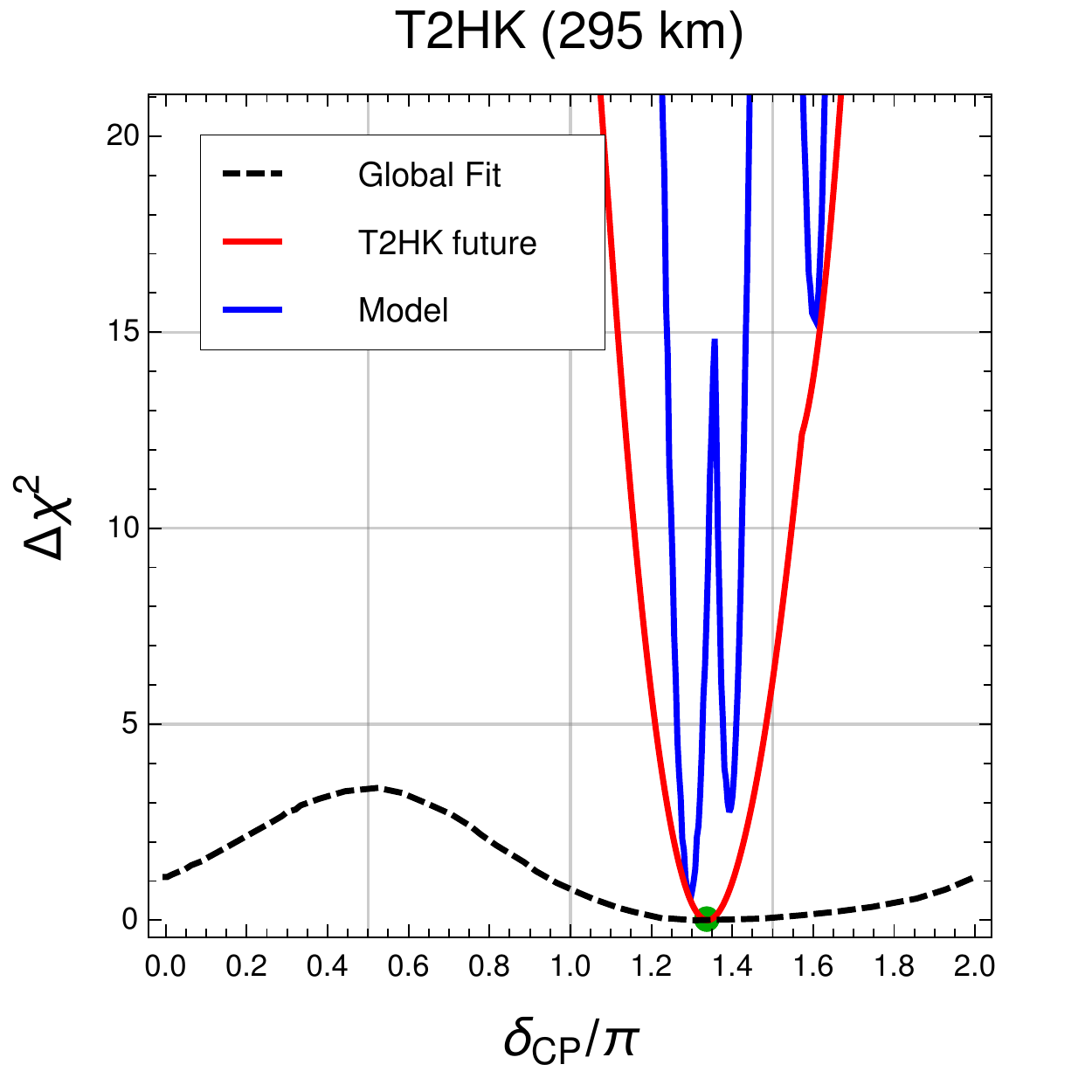}
  \caption{The left (right) panel indicates the reconstruction of
    oscillation parameters $\theta_{23}$ ($\delta_{CP}$). The green
    dot indicates the best fit value in the unconstrained oscillation
    picture, taken from~\cite{Forero:2014bxa}. The black dashed curve
    indicates the current global fit measurement, the red solid curve
    indicates the T2HK expectation for the measurement in the generic
    oscillation scheme, while blue solid curve represents the precise
    measurement by the model.}
 \label{octant_sensitivity2}
  \end{figure}

%%%%%%%%%%%%%%%%%%%%%%%

We now turn to the sensitivity of the future generation of planned
long baseline accelerator neutrino oscillation experiments such as
DUNE~\cite{Acciarri:2015uup} and T2HK~\cite{Abe:2011ts}, for
definiteness.
%
%The idea is to determine whether such experiments have the potential
%of ruling out our minimal benchmark oscillation theory.
%
Our results are depicted in
Figs.~\ref{octant_sensitivity1}~and~\ref{precision}.
The upper (lower) panel of Fig.\ref{octant_sensitivity1} corresponds to DUNE
(T2HK). Notice that in all plots of Fig.\ref{octant_sensitivity1}, there is an extra cyan band at
5$\sigma$ C.L.
One sees that they will have the potential of severely constraining
the parameter space of the model. 
The most important point to note is that they help to remove the
four-fold degeneracy to two-fold degeneracy, due to their fantastic
sensitivity to $\delta_{\rm CP}$. It excludes a large part of the
parameter space. The allowed region at 4$\sigma$ corresponds to the
1.10$\pi$ ($-162^\circ$) to 1.75$\pi$ ($-45^\circ$) for DUNE and for
maximal value of $\theta_{23}$, model predicts maximal CP violation
that $\delta_{CP}$ = $-90^\circ$. This is a very nice prediction of
the benchmark model~\cite{Chen:2015jta}. Notice that T2HK plays a
crucial role in removing the four-fold degeneracy of the CP phase
completely for most of the parameter space (for example, if
$\theta_{23}$ lies in the upper octant) and it improves the
sensitivity tremendously which can be attributed to the fact that T2HK
has very good sensitivity to the CP phase. For a fixed CP phase, it
also removes the octant degeneracy but not at 5$\sigma$ C.L. and that
can be easily verified by placing a horizontal line around the star
symbol on the left plot of the lower panel of
Fig.\ref{octant_sensitivity1}. Fig.~\ref{precision}
    displays the sensitivity region in $\delta_{CP}$ versus
    $\sin^2\theta_{23}$, clearly indicating the capability of T2HK
    (similar holds for DUNE) in establishing CP violation by rejecting
    the CP conservation scenario at more than 5$\sigma$ C.L. The
    figure gives a quantitative estimate of the precise
    ``measurement'' of $\sin^2\theta_{23}$ and $\delta_{CP}$ for the
    generic unconstrained 3-neutrino oscillation scenario, when NH is
    the true hierarchy. The star denotes the best-fit (unconstrained)
    values of the two parameters.  The true data have been generated
    with all the best-fit values of the oscillation parameters
    mentioned in sec.~\ref{simulation} and in the fit we have
    marginalized on solar and reactor mixing angles $\theta_{12}$ and
    $\theta_{13}$ respectively keeping NH fixed. The red, blue and
    dark green bands correspond to the 2$\sigma$, 3$\sigma$ and
    4$\sigma$ C.L uncertainty respectively at 1 d.o.f. Notice that in
    this case also the octant would remain unresolved even at
    2$\sigma$ C.L. 

  Before concluding let us also show the corresponding $\chi^2$
  profiles.  The plots in Fig~\ref{octant_sensitivity2} quantify the
  reconstruction capability for the oscillation parameters
  $\theta_{23}$ ($\delta_{CP}$). The green dot indicates the
  unconstrained best fit value from~\cite{Forero:2014bxa}. The black
  dashed curve indicates the current global fit measurement, while the
  red solid curve gives the T2HK expectation for the general
  oscillation scheme and the blue solid curve represents the precise
  measurement by the model.

\section{Summary and outlook :}

We have performed realistic simulations of the current long baseline
experiments T2K and NOvA as well as future ones such as DUNE and T2HK
in order to determine their potential in probing neutrino oscillation
parameters in general, as well as testing our ``minimal'' benchmark
theory of neutrino oscillations.
We have seen that the standard unconstrained three-neutrino picture
and our benchmark scenario predict different minima for the neutrino
oscillation parameters. Nevertheless, current neutrino oscillation
experiments cannot exclude our benchmark scenario.
In all our considerations we have had to assume a ``true'' value of
the oscillation parameters in order to determine the expected
precision of a future ``measurement''. This ``true'' value has been taken
from~\cite{Forero:2014bxa}.
However we could well have taken it from any of the other recent
global oscillation fits, namely those
in~\cite{Capozzi:2016rtj,Esteban:2016qun}.\\

 An obvious question arises, namely, what is the sensitivity of the
 model for any pair of unconstrained value of $\theta_{23}$ and
 $\delta_{CP}$? In other words, what are the values of $\theta_{23}$
 and $\delta_{CP}$ ``true'' for which the model can be confirmed or
 excluded at a given confidence?  With this in mind, we fix the true
 values of the currently ``best determined'' oscillation parameters
 $\Delta m_{ij}^2$, $\theta_{12}$ and $\theta_{13}$.
 Given their current errors their central values are not expected to
 change significantly in upcoming experiments.
 We now vary both $\theta^{\rm TRUE}_{23}$ and
 $\delta^{\rm TRUE}_{\rm CP}$, finding the corresponding minimum of
 $\chi^2$ within the benchmark scheme by varying the model parameters
 $\theta_\nu$ and $\phi_\nu$.
 This way we obtain a function
 $\chi_{\rm min}^2(\theta^{\rm TRUE}_{23},\delta^{\rm TRUE}_{\rm
   CP})$, 
\begin{equation}
\chi^2_{\rm min}(\theta^{\rm TRUE}_{23},\delta^{\rm TRUE}_{\rm CP})={\rm Min}[\chi^2_{\rm min}(\theta^{\rm TRUE}_{23},\delta^{\rm TRUE}_{\rm CP},\theta_\nu,\phi_\nu)\rightarrow{\rm over }\quad\theta_\nu,\phi_\nu]
\end{equation}
 \begin{figure}[h!]
\includegraphics[scale=.8]{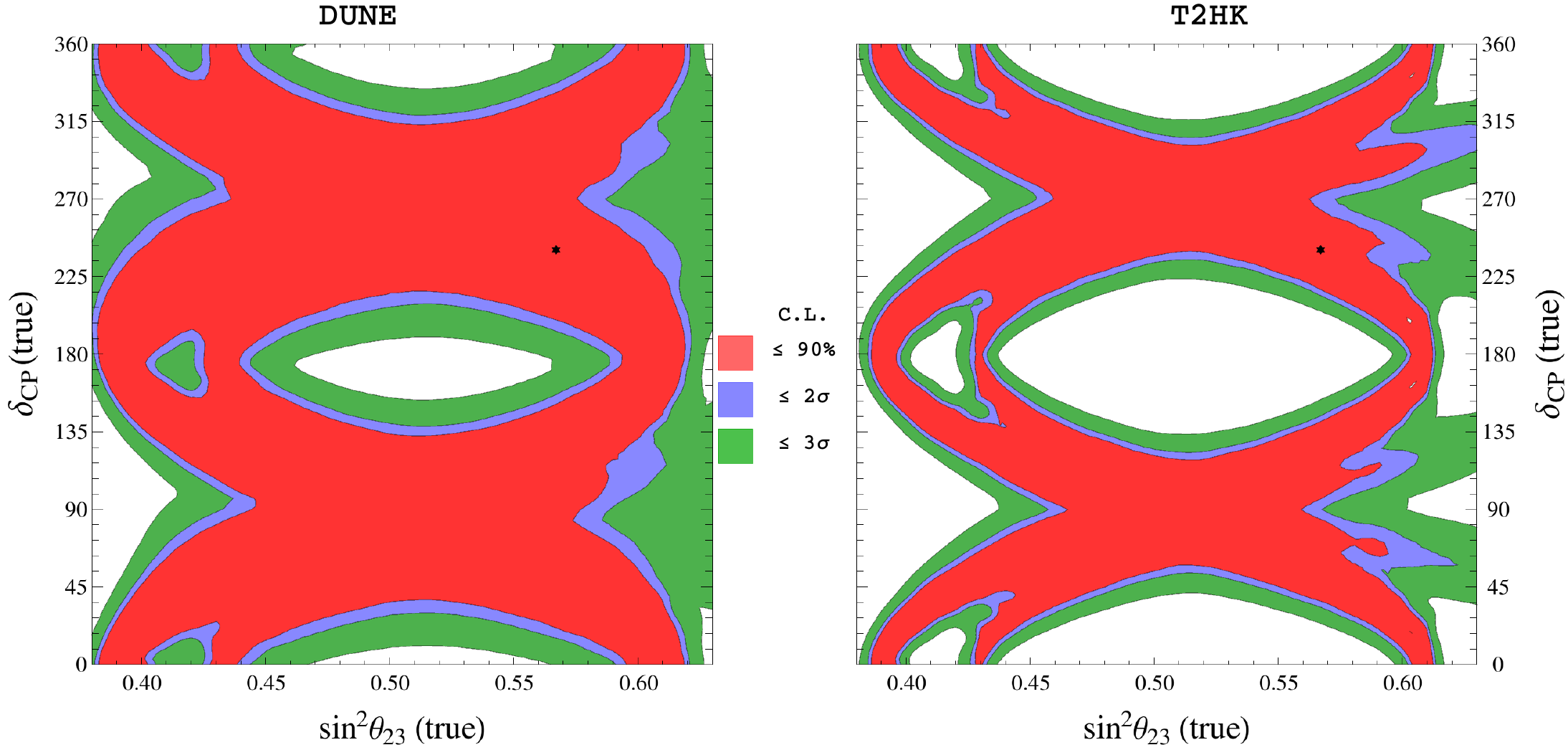}
\caption{Probing the model through the true values of
  $\sin^2\theta_{23}$ and $\delta_{CP}$ for normal neutrino mass
  ordering (NH). The shaded regions denote the confidence level at
  which DUNE (left) or T2HK (right) would confirm our minimal
  benchmark oscillation model. The red band corresponds to 90\%C.L.,
  the blue band corresponds to 2$\sigma$ C.L. and the dark green band
  corresponds to the 3$\sigma$ C.L. allowed region. The confidence
  levels are given for 1 d.o.f. ($\Delta\chi^2$ = 2.71, 4 and 9 respectively). The star denotes the unconstrained
  values taken from the fifth column of table \ref{tab:predic1}.}
 \label{region4}
\end{figure}

 Now for each true data set the new parameters are marginalized within
 their allowed values coming from Fig.~\ref{region1}. The resulting
 $\chi^2$ represents the ability of the experiment to probe the model
 if it measures a given value of $\theta^{\rm TRUE}_{23}$ and
 $\delta^{\rm TRUE}_{\rm CP}$ and it has been addressed very nicely in 
fig.~\ref{region4}.
%
%  In Fig.~\ref{region4} we present the confidence values with which any
%  pair of oscillation parameters $\sin^2\theta_{23}$ and $\delta_{CP}$
%  can be probed by the DUNE or T2HK setups, separately.
%
 The light red band corresponds to the 90\% C.L. region, the
 light blue band corresponds to 2$\sigma$ C.L. region and the green
 band corresponds to the 3$\sigma$ C.L region. 
 The blank region indicates the unconstrained parameter space of
 $\theta_{23}$ and $\delta_{CP}$ for which the model can be excluded
 at more than 3$\sigma$ C.L..
 In short, our ``minimal'' benchmark oscillation model serves to
 highlight the increased sensitivity of the new planned future generation of
 long baseline oscillation experiments.\\

\section*{Acknowledgments}

This research is supported by the Spanish grants FPA2014-58183-P,
Multidark CSD2009-00064, SEV-2014-0398 (MINECO) and
PROMETEOII/2014/084 (Generalitat Valenciana). P. S. P. acknowledges the support of
FAPESP grant 2014/05133-1, 2015/16809-9 and 2014/19164-6.

 % \begin{appendix}
  
 % \end{appendix}

 \bibliographystyle{naturemag}
 \bibliography{merged_Valle,newrefs,d27}  
 
\end{document}